\documentclass[letterpaper,11pt]{article}
\pdfoutput=1 

\usepackage{jcappub} 

\usepackage[T1]{fontenc} 

\def\beq{\begin{equation}}
\def\eeq{\end{equation}}

\title{Inflationary models constrained by reheating}

\author[]{Gabriel Germ\'an,}
\author[]{Juan Carlos Hidalgo,}
\author[]{Luis E. Padilla}

\affiliation[]{Instituto de Ciencias F\'{\i}sicas, Universidad Nacional
Aut\'onoma de M\'exico,\\ Av. Universidad s/n, Cuernavaca, Morelos, 62210, Mexico}

\emailAdd{gabriel@icf.unam.mx}
\emailAdd{hidalgo@icf.unam.mx}
\emailAdd{padilla@icf.unam.mx}

\abstract{The study of reheating in inflationary models is crucial to understand the early universe and obtain information about the dynamics and parameters of inflation. The reheating temperature $T_{re}$ and the duration of the reheating phase, quantified by the number of $e$-folds $N_{re}$, have significant implications for particle production, thermalisation, and the primordial power spectrum. The duration of reheating affects the abundance of particles, including dark matter, and shapes the primordial power spectrum and the anisotropies of the cosmic microwave background. By combining cosmological observations and theoretical considerations, we can constrain both $T_{re}$ and $N_{re}$, which in turn restrict the spectral index $n_s$, the tensor-to-scalar ratio $r$, and the parameters of the inflationary model. The use of consistency relations between observables, such as $n_s$ and $r$, provides additional constraints on inflationary models and determines limits for other observables, such as the running of the scalar spectral index. These limits are valuable for assessing the viability of models and can serve to specify priors in Bayesian analyses of specific models. As an example of how to proceed, we study in detail a particular case of a generalized $\alpha$-attractor model that accurately reproduces the observed quantities. We present equations for the conditions of instantaneous reheating, establish consistency relations, and explore the generalized $\alpha$-attractor model using cosmological data. We study the model defined by its potential as a given formula and, furthermore, considering its possible origin in supergravity theories, we impose constraints on the reheating temperature to avoid the overproduction of gravitinos.}

\begin{document}
\maketitle
\flushbottom

\section{Introduction}\label{int}

Reheating in inflationary models plays a crucial role in refining our understanding of the early universe and provides valuable insights into the dynamics and parameters of inflationary scenarios. The reheating temperature, denoted as $T_{re}$, represents the maximum temperature reached during reheating and influences particle production, thermalisation, and the primordial power spectrum. Constrained estimates of the reheating temperature involve a combination of cosmological observations, particularly those derived from the cosmic microwave background radiation, and theoretical considerations(for reviews on inflation, see e.g., \cite{Linde:1984ir}-\cite{Odintsov:2023weg}; for reviews on reheating, see e.g.,  \cite{Bassett:2005xm}-\cite{Amin:2014eta}).

The duration of the reheating phase represents the transition from the inflationary era to a radiation-dominated era. It directly impacts particle production, thermalisation, and the abundance of particles, including relic particles such as dark matter. The duration of reheating, quantified by the number of $e$-folds $N_{re}$, also affects the generation and amplitude of primordial perturbations, shaping predictions for the primordial power spectrum and observable anisotropies in the cosmic microwave background radiation.
When properly constrained, these two aspects of inflationary cosmology can provide constraints on observables through consistency relations. These relations involve quantities such as the spectral index $n_s$, the tensor-to-scalar ratio $r$, and the running of the scalar spectral index $dn_s/d\ln k$. They offer additional constraints on inflationary models and arise from the dynamics of the inflationary field, allowing for tests of internal consistency within cosmological models. Deviations from these relations provide valuable insights, indicating deviations from model assumptions or the presence of additional physical effects.

These aspects of inflationary cosmology are studied here independently of the model and are applied to an important class of models known as $\alpha$-attractor inflation models. These models provide a framework for studying inflationary physics, drawing inspiration from concepts in supergravity and string theory and showing intriguing connections with hyperbolic or logarithmic geometries in field space. $\alpha$-attractor models accurately reproduce observed quantities and exhibit an attractor behavior, expanding the range of initial conditions that lead to similar observational outcomes.

The paper is organized as follows: In Section $\ref{model}$, the conditions for instantaneous reheating are studied, deriving an equation that allows determining the values of the spectral index and the tensor-to-scalar ratio for which $N_{re}=0$, along with the maximum number of inflationary and radiation $e$-folds. Consistency relations between observables are established, including the spectral index, tensor-to-scalar ratio, and the running of the scalar spectral index, providing immediate constraints on these observables. Additionally, a formula for the instantaneous reheating temperature is obtained. In Section $\ref{alfa}$, a generalization of the basic $\alpha$-attractor model is presented, ensuring a positive-definite potential, making it viable for odd and fractional values of the exponent $p$. In particular, this class of potentials exhibits quadratic behavior around its minimum for any value of $p$. Subsection $\ref{caso1}$ focuses on the case of $p = 2$, including running, within the cosmological model $\Lambda$CDM$+r+dn_s/d\ln k$, using data from Planck TT, TE, EE + lowE + lensing + BK15 + BAO. In Subsection $\ref{caso2}$, the same case of $p = 2$ is studied, but without running, in the cosmological model $\Lambda$CDM$+r$ using data from the Planck and BICEP/Keck 2018 collaborations. In Subsection $\ref{caso3}$, the dependence of cosmological quantities on the parameter $\alpha$ is investigated. When considering the possible origin of the model in supergravity theories, restrictions on the reheating temperature must be imposed to avoid overproduction of gravitinos. Therefore, in Subsection $\ref{caso4}$, the gravitino problem is considered, which requires lower reheating temperatures. Our conclusions are presented in Section $\ref{con}$.

\section{The condition of instantaneous reheating}\label{model}
In this section, we study the conditions for instantaneous reheating, which should help determine wide bounds on cosmological quantities of interest. As we will see, instantaneous reheating is equivalent to zero e-folds of expansion during reheating, i.e., $N_{re}=0$. This is an idealized situation because mechanisms such as broad parametric resonance  \cite{Mukhanov:2005sc} or instant preheating \cite{Felder:1998vq} can significantly contribute to $N_{re}$. However, it should be clear that this idealized situation is important because it provides broad bounds that can then be used in Bayesian analyses with a high level of confidence that a range of parameter values is not being disregarded, thus allowing for increased reliability in the analysis.

 Various reheating mechanisms have been extensively discussed in the literature and play a crucial role in the dynamics of the inflaton field. Broad parametric resonance, involve oscillations of the inflaton field around the minimum of the potential and result in the generation of a significant number of particles, effectively reheating the universe after approximately 20 oscillations. Another mechanism, instant preheating, occurs when the inflaton is close to the bottom of its potential. Initially low-mass particles are created, but after approximately half an oscillation, they acquire a much higher mass, reaching the scale of the Grand Unified Theory (GUT). These particles then decay into heavy fermions through a Yukawa interaction, which further decay into lighter particles, forming a plasma of relativistic fluid. With subsequent oscillations of the inflaton, this plasma becomes dominant, leading to the reheating of the universe.

 In classical reheating, the energy of the inflaton field is converted into particles and radiation in a thermal manner. After inflation, the inflaton field oscillates around the minimum of its potential and decays into elementary particles. These particles follow thermal distributions, either Bose-Einstein or Fermi-Dirac, depending on their type, and their production occurs through scattering and annihilation processes. This transfers energy to the plasma of the early universe. The equation-of-state parameter (EoS) during reheating, denoted as $\omega_{re}$, characterizes this epoch. The EoS is not constant but changes from -1/3 at the end of inflation to 1/3 at the beginning of radiation domination. However, considering the effective equation-of-state parameter provides an average value during reheating, which remains constant.

 This study focuses on exploring the constraints imposed by reheating on inflationary models. The main concept examined is instantaneous reheating, where there are precisely zero e-folds from the end of inflation to the beginning of the radiation epoch. Although this restriction may initially appear strict, the resulting constraints are crucial for evaluating the feasibility of models. These constraints also serve as prior specifications in Bayesian analyses of specific models as has been done in e.g., \cite{Cedeno:2019cgr}, \cite{Iacconi:2023mnw}, \cite{LinaresCedeno:2023tcb}.

As an illustrative example, we consider the basic $\alpha$-attractor model. This model features a single parameter (excluding the overall scale $V_0$) and exhibits consistency relations connecting the spectral index $n_s$, the $\alpha$-parameter, and the tensor-to-scalar ratio $r$. In more complex models involving additional parameters, it is theoretically possible to derive consistency relations that incorporate those parameters as well. However, in cases where the equations become too complex to obtain analytical solutions, numerical methods can be employed to follow the outlined procedure.

\begin{figure}[ht!]
\centering
\includegraphics[trim = 0mm  0mm 1mm 1mm, clip, width=13cm, height=8cm]{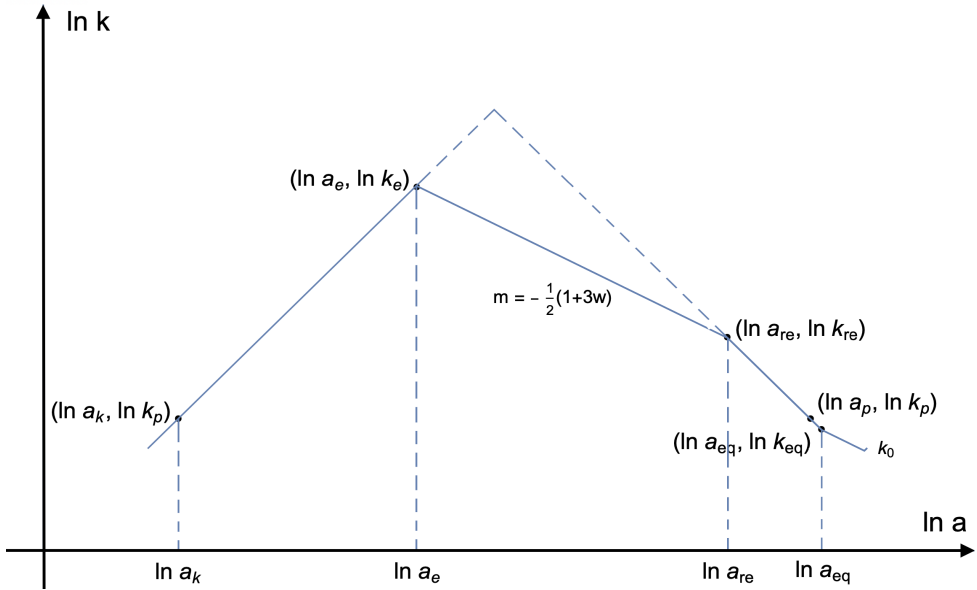}
\caption{\small The diagram provides a representation of the universe's evolution, illustrating the logarithm of the comoving Hubble scale wavenumber mode $\ln k$ (where $k=a H$) as a function of the logarithm of the scale factor $\ln a$. During inflation, the comoving scale wavenumber $k$ exits the horizon, and it reenters when the scale factor is $a_p < a_{eq}$, with $a_{eq}$ denoting the scale factor at radiation-matter equality.
The reheating phase is depicted by a line with an arbitrary slope of $m=-\frac{1}{2}\left(1+3\omega_{re}\right)$, where $\omega_{re}$ is an assumed constant effective equation of state parameter (EoS). The dashed lines can be used to extend the inflation and radiation lines to the vertex (by shifting the reheating line parallel to itself), representing instantaneous reheating with a vanishing number of $e$-folds of reheating. Reheating lines with $\omega_{re}>1/3$ ($m < -1$) are positioned to the right of the dashed radiation line (not shown).
The diagram also emphasizes that the number of $e$-folds $N_{keq}$ from the moment when scales leave the horizon during inflation to the time of radiation-matter equality is equal to the sum of the number of $e$-folds during inflation $N_k$, reheating $N_{re}$, and radiation $N_{rd}$, regardless of the specific characteristics and durations of each phase individually, as indicated by Eq.~(\ref{Nkeq}). The figure also suggests a general trend: as the duration of reheating decreases, the durations of the inflationary and radiation epochs increase, reaching their maximum values when $N_{re}=0$.}
\label{fd}
\end{figure}

The number of $e$-folds during reheating, denoted by $N_{re}$, is commonly expressed in terms of the logarithm of scale factors. Alternatively, it can be defined in relation to the energy densities at the end of inflation and at the end of reheating, as follows
\begin{equation}
N_{re}= \ln\left(\frac{a_{re}}{a_{e}}\right)=\frac{1}{3(1+\omega_{re})} \ln\left(\frac{\rho_e}{\rho_{re}}\right).
\label{Nre1}
\end{equation}
The second equality can be derived from the solution $\rho\propto a^{-3(1+\omega_{re})}$ of the fluid equation, assuming a constant effective equation of state (EoS) parameter $\omega_{re}$. In order to determine the value of $a_e$, we consider the number of $e$-folds from the moment when the scale factor was $a_k$, corresponding to the time when the pivot scale with wavenumber $k_p=0.05/$Mpc exited the horizon, to the time of radiation-matter equality at $a_{eq}$ \cite{German:2020iwg}
\begin{equation}
N_{keq}\equiv \ln\left(\frac{a_{eq}}{a_{k}}\right)= \ln\left(\frac{a_{eq} H_k}{a_{k}H_k
}\right)=\ln\left(\frac{a_{eq} H_k}{k_p}\right)=\ln\left(\frac{a_{eq} \pi \sqrt{A_s}M_{Pl}}{\sqrt{2}\,k_p}\sqrt{r}\right),
\label{Nkeq}
\end{equation}
where $k\equiv a_kH_k=k_p$, and the equation for the amplitude of scalar perturbations was used in the final equality. It is noteworthy that this equation, which can be expressed as the sum of the number of $e$-folds during inflation $N_k$, reheating $N_{re}$, and radiation $N_{rd}$, remains independent of the specific characteristics and durations of each individual epoch. This is evident from the Friedmann diagram depicted in Fig.~\ref{fd}, where the sum of the projections along the $\ln a$ axis is the relevant quantity for Eq.~(\ref{Nkeq}), rather than the individual lengths and characteristics of each projection. This property holds true even when the approximation of straight lines in Fig.~\ref{fd} is replaced by more realistic curves.

Because the number of $e$-folds of inflation is given by $N_k\equiv \ln\left(\frac{a_{e}}{a_{k}}\right)$, from Eq.~(\ref{Nkeq}) we get
\begin{equation}
a_{k}=\frac{\sqrt{2}\,k_p}{\pi\sqrt{A_s} M_{Pl}\sqrt{r}},  \quad\quad  a_{e}=\frac{\sqrt{2}\,k_pe^{N_k}}{\pi\sqrt{A_s} M_{Pl}\sqrt{r}}.
\label{ae}
\end{equation}
From the first equation, we can understand how fixing $r$ determines the number of $e$-folds from horizon crossing during inflation to the time of radiation-matter equality since fixing $r$ is equivalent to fixing $a_k$.

The number of $e$-folds during the radiation epoch is determined from the reasonable assumption of entropy conservation following reheating. At the end of reheating, the energy density in terms of the temperature  is given by $\rho_{re}=\frac{\pi^2 g_{re}}{30}T_{re}^4$, where $g_{re}$ represents the effective number of relativistic species. Entropy conservation establishes a relationship between the reheating entropy, preserved in the cosmic microwave background (CMB), and the present neutrino background
\begin{equation}
g_{s,re} T_{re}^3 =  \left(2T_0^3+6\times\frac{7}{8}T_{\nu 0}^3\right) \left(\frac{a_0}{a_{re}}\right)^3.
\label{gsre}
\end{equation}
The effective number of species for entropy conservation is denoted as $g_{s,re}$. The present temperature of neutrinos is given by $T_{\nu 0}=\left(\frac{4}{11}\right)^{1/3} T_0$. This establishes a relationship between the reheating temperature and the current temperature of the CMB, which can be expressed as follows
\begin{equation}
T_{re} = T_0\left(\frac{43}{11g_{s,re}}\right)^{1/3} \frac{a_0}{a_{eq}}\frac{a_{eq}}{a_{re}} = T_0\left(\frac{43}{11g_{s,re}}\right)^{1/3} \frac{a_0}{a_{eq}}e^{N_{rd}}, 
\label{Tre1}
\end{equation}
where $N_{rd}$ is the number of $e$-folds during radiation
\begin{equation}
N_{rd}\equiv \ln\left(\frac{a_{eq}}{a_{re}}\right)=\ln\left(\frac{a_{eq}T_{re}}{\left(\frac{43}{11g_{s,re}}\right)^{1/3}a_0T_0}\right).
\label{Nrd1}
\end{equation}
Here, $a_{re}$ is the scale factor at the end of reheating which can be expressed as (we set $a_0 = 1$ for convenience)
\begin{equation}
a_{re}=\left(\frac{43}{11g_{s,re}}\right)^{1/3}\frac{T_0}{T_{re}}.
\label{are}
\end{equation}
From the first equality in Eq.~(\ref{Nre1}) we have
\begin{equation}
N_{re}=\ln\left(\frac{\left(\frac{43}{11g_{s,re}}\right)^{1/3}\sqrt{A_s\,r}\,\pi M_{Pl}T_0\,e^{-N_k}}{\sqrt{2}\,k_pT_{re}}\right).
\label{Nre2}
\end{equation}
On the other hand, the energy densities at the end of inflation and at the end of reheating can be expressed as 
\begin{equation}
\rho_{e}=\frac{3}{2}V_e,   \quad\quad \rho_{re}=\frac{\pi^2 g_{re}}{30}T_{re}^4,
\label{ros}
\end{equation}
respectively. Here, $V_e$ is the inflationary potential at the end of inflation. Without loss of generality, we can express the potential as $V(\phi)=V_0 f(\phi)$, where $V_0$ is the overall scale and $f(\phi)$ the rest of the potential containing all the terms dependent on $\phi$. By writing $V_e=\frac{V_e}{V_k}V_k= 3\frac{V_e}{V_k}H_k^2M_{Pl}^2$, we obtain
\begin{equation}
V_e=\frac{3}{2}\pi^2A_s\, r \frac{f(\phi_e)}{f(\phi_k)} M_{Pl}^4.
\label{Ve}
\end{equation}
Here, $\phi_k$ ($\phi_e$) is the value of the inflaton at horizon crossing (end of inflation) and $A_s$ is the amplitude of scalar perturbations.
By substituting (\ref{ros}), (\ref{Ve}) into (\ref{Nre1}) we can derive an alternative expression for $N_{re}$
\begin{equation}
N_{re}=\frac{1}{3(1+\omega_{re})} \ln\left(\frac{135 A_s\, r M_{Pl}^4 f(\phi_e)}   {2 g_{re} T_{re}^4 f(\phi_k)}\right).
\label{Nre4}
\end{equation}
It is worth noting that Eq.~(\ref{Nre4}) does not exhibit any singularities with respect to $\omega_{re}$. However, this formula has limited applicability as it only provides broad bounds for $N_{re}$ within a particular model. To determine these bounds more precisely, it is advantageous to maximize $N_{re}$ by considering the minimum reheat temperature for a given $r$\footnote{for a model-independent approach see  \cite{German:2022sjd}.}.
From Eqs.~(\ref{Nre2}) and (\ref{Nre4})  we get the reheat temperature
\begin{equation}
T_{re}= ln\left(\frac{\sqrt{2}\,k_p\,e^{N_k}}{\left(\frac{43}{11g_{s,re}}\right)^{1/3}\sqrt{A_s\,r}\,\pi T_0}\right)^{\frac{3(1+\omega_{re})}{1-3\,\omega_{re}}}  \left(\frac{135 A_s\, r  f(\phi_e)}   {2 g_{re}  f(\phi_k)}\right)^{\frac{1}{1-3\,\omega_{re}}} M_{Pl},
\label{Tre}
\end{equation}
an expression valid for $\omega_{re}\neq 1/3$. To find  $N_k$, $f(\phi_e)$, and $f(\phi_k)$, the values of $\phi_e$ and $\phi_k$ are required. The equation $r = 16\epsilon$ at the time of horizon crossing can be solved to find $\phi_k$. Thus, it is always possible to express $T_{re}$ as a function of $r$, the model parameters, and the equation of state $\omega_{re}$.
Substituting $T_{re}$ as given by Eq.~(\ref{Tre}) into (\ref{Nre2}) or (\ref{Nre4}) we find
\begin{equation}
N_{re}=\frac{1}{1-3\,\omega_{re}}\ln\left( \frac{\left(\frac{43}{11g_{s,re}}\right)^{4/3}g_{re}A_s\pi^4T_0^4e^{-4N_k}f(\phi_k)r}{270k_p^4\,f(\phi_e)}\right).
\label{Nre5}
\end{equation}
This equation, although derived in a very simple but different way, and applicable to any potential $V(\phi)=V_0f(\phi)$, is equivalent to the one investigated in \cite{Martin:2013tda}, \cite{Dai:2014jja}, \cite{Munoz:2014eqa},  \cite{Cook:2015vqa}, and in several subsequent papers referencing their work (for a sample of papers with emphasis on $\alpha$-attractors see \cite{Odintsov:2016vzz} -\cite{Drewes:2023bbs} and on reheating \cite{Khlopov:1984pf} - \cite{Haque:2021dha}). 
The condition for instant reheating ($a_{re} = a_e$), which implies $N_{re} = 0$, can be expressed as follows
\begin{equation}
N_k=S_1+\frac{1}{4}\ln\left(\frac{f(\phi_k) r}{f(\phi_e)}\right),
\label{cir}
\end{equation}
where 
\begin{equation}
S_1 = \frac{1}{4}\ln\left(\frac{\left(\frac{43}{11g_{s,re}}\right)^{4/3}g_{re}\pi^4 A_s T_0^4 }{270k_p^4 }\right)\approx 56.7.
\label{Tc}
\end{equation}
In addition to $N_k$ the second term in the rhs of Eq.~(\ref{cir})  includes all the model dependent parts and the first term, $S_1$, is approximately equal to 56.7 for $g_{s,re}=g_{re}=106.75$. 

Thus, for a given model of inflation we calculate $N_k$ in terms of $r$ and the parameters of the model, denoted $\alpha$. The solution of Eq.~(\ref{cir})  gives the parameters at instant reheating, $\alpha_{ir}$, which guarantees a lower bound, $N_{re}=0$, for the number of $e$-folds during reheating and an upper bounds for $N_k$ and $N_{rd}$. From the consistency relations of the model we determine all the relevant observables.

The condition (\ref{cir}) can alternatively be derived by equating (\ref{Tre}) to itself for two different values of $\omega_{re}$. This is because in the $n_s$-$T_{re}$ plane, all curves converge to a single point representing the instant reheating temperature (see e.g., Fig.~\ref{Treir} of Section \ref{alfa}). Solving this equation in a specific model provides the value of the model parameters $\alpha_{ir}$, which corresponds to the point of instant reheating, for a given $r$. 

The obtained value for $\alpha_{ir}$ represents the maximum or instantaneous reheat temperature. It also determines the maximum number of $e$-folds of radiation and, in the case of $\omega_{re} < \frac{1}{3}$, the maximum number of $e$-folds of inflation. However, if $\omega_{re} > \frac{1}{3}$, this bound becomes a minimum number of $e$-folds of inflation. This behavior can be easily understood by referring to the diagram depicted in Fig.~\ref{fd}. For a comprehensive discussion, refer to \cite{German:2022sjd}.

 It is also possible to directly obtain the value of the spectral index $n_s$ using the consistency relation to eliminate the parameter $\alpha$. This is demonstrated in the example of Section \ref{alfa}.

In a given inflationary model, $\alpha_{ir}$ can provide more stringent bounds on observables compared to experimental constraints such as Planck/Keck bounds. Once the value of $\alpha_{ir}$ is determined, bounds for all observables can be obtained through the consistency relations of the model, as well as the number of $e$-folds of inflation, reheating, and radiation, as demonstrated in the example (see also  \cite{Garcia:2023tkk}). These bounds are valuable for assessing the viability of models and can serve as priors in Bayesian analyses of specific models.

A concise expression for the instantaneous reheating temperature can be derived from Eq.~(\ref{Nre4})
\begin{equation}
T_{re}^{(ir)}=\left.\left(\frac{135 A_s\, r f(\phi_e)} {2 g_{re} f(\phi_k)}\right)^{\frac{1}{4}}\right\rvert_{\alpha=\alpha_{ir}} M_{Pl}.
\label{Treir1}
\end{equation}
This temperature also represents the maximum reheat temperature for a given $r$ and, notably, it is independent of $\omega_{re}$.

It is important to highlight that by adding an undetermined $N_k$ to $N_{re}$ from Eq. (\ref{Nre5}) and to $N_{rd}$ from Eq. (\ref{Nrd1}), where $T_{re}$ is determined by Eq. (\ref{Tre}), we precisely arrive at Eq. (\ref{Nkeq}). Remarkably, this equation was derived using a straightforward approach that solely relied on the slow-roll (SR) approximation to express $H_k$ from the equation for the scalar amplitude of density perturbations, Eq. (\ref{IA}).
In a Friedmann diagram (shown in Fig. \ref{fd}), where the logarithm of the scale factor $\ln a$ is plotted against the logarithm of the wavenumber mode $\ln k$, we can clearly observe that $N_{keq} \equiv \ln \frac{a_{eq}}{a_k} = N_k + N_{re} + N_{rd}$ regardless of the specific characteristics and durations of inflation, reheating, and radiation individually. Hence, we can use Eq. (\ref{Nkeq}) to verify the accuracy of our results solely by knowing the corresponding value of $r$ in a given calculation, as demonstrated in Section \ref{alfa} and summarized in the Tables \ref{tabla} and \ref{tablanr}.

Finally, by combining Eqs.~(\ref{Nre2}) and (\ref{Nre4}), we can obtain an expression for $\omega_{re}$ in terms of $n_s$ as follows:
\begin{equation}
\omega_{re}=-1-\frac{1}{3}\frac{\ln\left(\frac{135A_sM_{Pl}^4 f(\phi_e)r}{2g_{re}T_{re}^4f(\phi_k)}\right)}{N_k-\ln\left(\frac{\left(\frac{43}{11g_{s,re}}\right)^{1/3}\pi\sqrt{A_s}M_{Pl}T_0\sqrt{r}}{\sqrt{2}\,k_pT_{re}}\right)},
\label{wns}
\end{equation}
where the dependence on the spectral index $n_s$ arises from a consistency relation specific to the given model, which relates $r$ to $n_s$ and $\alpha$. This formula is further illustrated in Figs.\ref{wnsTre} and \ref{wnsAlfa} in Section~\ref{caso1}.

\section{A class of $\alpha$-attractor models}\label{alfa}

In this section we start by considering a class of models characterized by the potential
\begin{equation}
V = V_0\left(1-sech^{p}\left(\frac{\phi}{\sqrt{6\alpha}M_{Pl}}\right)\right),
\label{potsech}
\end{equation}
where $p$ is a positive real number. The potential $(\ref{potsech})$ can be seen as a generalization of the basic potential introduced in \cite{Kallosh:2013yoa}
\begin{equation}
V = V_0\tanh^2\left(\frac{\phi}{\sqrt{6\alpha}M_{Pl}}\right) = V_0\left(1-sech^{2}\left(\frac{\phi}{\sqrt{6\alpha}M_{Pl}}\right)\right).
\label{potanh}
\end{equation}
This generalization, as discussed in \cite{German:2021rin}, maintains simplicity while ensuring that the potential is positive-definite, making it viable for any reasonable value of $p$, including odd and fractional values. An interesting characteristic of the potential \eqref{potsech} is that it exhibits quadratic behaviour around its minimum for any value of $p$. In the vicinity of the minimum, the potential \eqref{potsech} can be approximated as
\begin{equation}
\frac{V}{V_0} = \frac{1}{2}p \left(\frac{\phi}{\sqrt{6\alpha}M_{Pl}}\right)^2 - \frac{1}{24}p(2+3p)\left(\frac{\phi}{\sqrt{6\alpha}M_{Pl}}\right)^4 + \dots, 
\label{potorigins}
\end{equation}
where $\alpha$ represents a dimensionless parameter. In the following subsections we study the case $p=2$, where $\omega_{re}=0$. This case has been partially studied  in several papers  \cite{German:2021rin}, \cite{German:2020cbw}, \cite{Iacconi:2023mnw} . Here, we give a complete an unified description including Planck bounds for the case $\Lambda$CDM+$r$+$ \frac{d n_s}{d \ln k}$ (with running) \cite{Akrami:2018odb} and without running \cite{Paoletti:2022anb}.

Models of inflation can be associated with cosmological observables. In the first-order approximation of the SR scenario, these observables can be expressed as follows (see, for example, \cite{Lyth:1998xn} and \cite{Liddle:1994dx})
\begin{eqnarray}
n_{t} &=&-2\epsilon = -\frac{r}{8} , \label{Int} \\
n_{s} &=&1+2\eta -6\epsilon ,  \label{Ins} \\
n_{sk}\equiv \frac{d n_s}{d \ln k} &=&16\epsilon \eta -24\epsilon ^{2}-2\xi_2, \label{Insk} \\
A_s &=&\frac{1}{24\pi ^{2}} \frac{V}{\epsilon\, M_{Pl}^{4}}. \label{IA} 
\end{eqnarray}
We use the following notation: $M_{Pl}=2.43568\times 10^{18} \mathrm{GeV}$ represents the reduced Planck mass. We have $n_t$ as the tensor spectral index, $r$ as the tensor-to-scalar ratio, and $n_s$ as the scalar spectral index. The running of the scalar index is $n_{sk}$  (often denoted as $\alpha$). The amplitude of density perturbations at a specific wavenumber $k$ is denoted by $A_s$. All these quantities are evaluated at the moment of horizon crossing for a wavenumber of $k_p=0.05$/Mpc. The SR parameters involved in the aforementioned expressions are as follows
\begin{equation}
\epsilon \equiv \frac{M_{Pl}^{2}}{2}\left( \frac{V^{\prime }}{V }\right) ^{2},\quad
\eta \equiv M_{Pl}^{2}\frac{V^{\prime \prime }}{V}, \quad
\xi_2 \equiv M_{Pl}^{4}\frac{V^{\prime }V^{\prime \prime \prime }}{V^{2}},
\label{Spa}
\end{equation}
where primes on $V$ denote derivatives with respect to the inflaton $\phi$.

\subsection {\bf The case $p=2$ with running}\label{caso1} 

We consider the bounds provided by the Table 3 of \cite{Akrami:2018odb} for the cosmological model $\Lambda$CDM$+r+dn_s/d\ln k$ with the data set Planck TT,TE,EE+lowE+lensing +BK15+BAO. These bounds offer constraints on the parameters and observables within the specific cosmological model and data combination
\begin{equation}
n_s =0.9658\pm 0.0040    \quad    (68\%\,\, C.L.),
\label{boundsns} 
\end{equation}
\begin{equation}
r < 0.068     \quad    (95\%\,\, C.L.).
\label{boundcr} 
\end{equation}

\begin{figure}[ht!]
\centering
\includegraphics[trim = 0mm  0mm 1mm 1mm, clip, width=12cm, height=8cm]{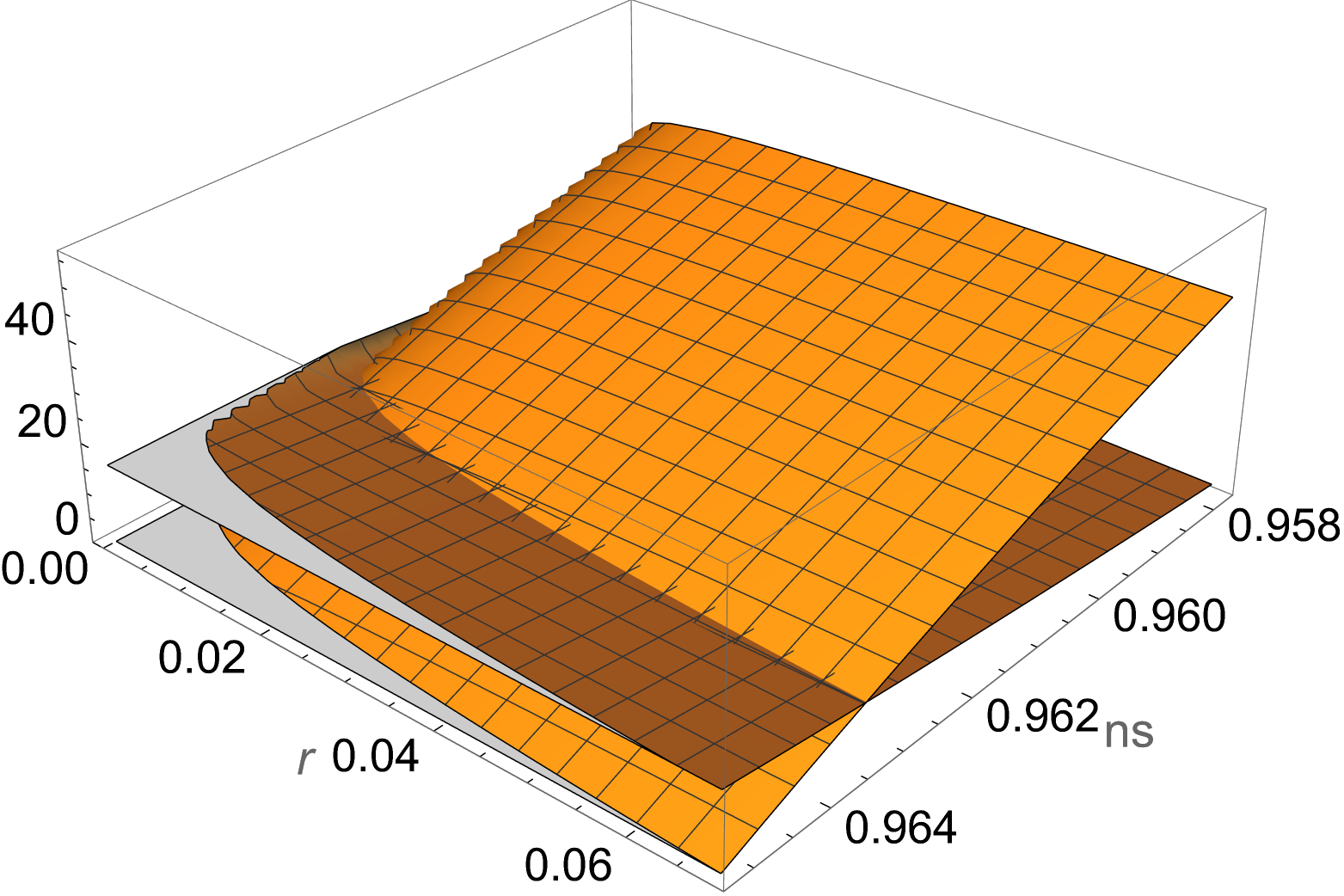}
\caption{The plot illustrates the relationship between the number of $e$-folds during reheating ($N_{re}$) represented by an orange surface and the logarithm of the reheat temperature ($\log_{10} T_{re}$) depicted as a brown surface. These quantities are plotted as functions of $n_s$ and $r$ for the $p=2$ model, assuming an equation of state parameter of $\omega_{re}=0$. To avoid cluttering the figure, the graphs for $N_k$ and $N_{rd}$ are not included, although they would fit equally well. At $2\sigma$ confidence level  the original range for the scalar spectral index is $0.9578 < n_s < 0.9738$ \cite{Akrami:2018odb}, and the tensor-to-scalar ratio satisfies $r < 0.068$.
However, the condition of instant reheating, $N_{re}= 0$, further restricts the allowed values of $n_s$ and $r$ to the ranges $0.9578<n_s<0.9652$ and $7.5 \times 10^{-17} < r < 0.068$, respectively. From these bounds, we can derive new constraints for the number of $e$-folds during inflation $N_{k}$, radiation $N_{rd}$, and the reheating temperature $T_{re}$. From the consistency relations of the models we obtain bounds for tensor index $n_t$ and the running index $n_{sk}$, as presented in the Table~\ref{tabla}.
It is worth noting that the number of $e$-folds during reheating strongly depends on the spectral index $n_s$ but not as much on the tensor-to-scalar ratio $r$. Even if we had precise knowledge of the value of $r$, a range of values in $n_s$ as depicted in the figure would result in an uncertainty of approximately 40 $e$-folds during reheating. Similar considerations can be made for the other cosmological quantities mentioned above.}
\label{Nrebounds}
\end{figure}

The key quantities required to solve Eq.~(\ref{cir}) are the values of the inflaton field at horizon crossing, denoted $\phi_k$, and at the end of inflation, $\phi_e$.
To express $\phi_k$ solely in terms of $r$ and the model parameter $\alpha$, we solve the equation $r=16\epsilon$ at horizon crossing, resulting in
\begin{equation}
\cosh^2\left(\frac{\phi_k}{\sqrt{6\alpha}M_{Pl}}\right) = \frac{32}{\sqrt{3r\alpha(64+3r\alpha)}-3r\alpha}.
\label{fik}
\end{equation}
By setting $\epsilon=1$, we obtain
\begin{equation}
\cosh^2\left(\frac{\phi_e}{\sqrt{6\alpha}M_{Pl}}\right) = \frac{2}{\sqrt{3\alpha(4+3\alpha)}-3\alpha}.
\label{fie}
\end{equation}
In terms of the potential function $f(\phi)$ defined as $V(\phi) = V_0 f(\phi)$, we have
\begin{equation}
f(\phi_k) = 1 - \frac{\sqrt{3r\alpha(64+3r\alpha)}-3r\alpha}{32},
\label{fk}
\end{equation}
\begin{equation}
f(\phi_e) = 1 - \frac{\sqrt{3\alpha(4+3\alpha)}-3\alpha}{2}.
\label{fe}
\end{equation}
The number of $e$-folds during inflation, $N_{k} = -\frac{1}{M_{Pl}^2}\int_{\phi_k}^{\phi_e}\frac{V}{V'}d\phi$, is given by
\begin{equation}
N_k = \frac{3}{2}\alpha\left(\cosh\left(\frac{\phi_k}{\sqrt{6\alpha}M_{Pl}}\right)^2 - \cosh\left(\frac{\phi_e}{\sqrt{6\alpha}M_{Pl}}\right)^2\right),
\label{Nk0}
\end{equation}
combining with Eqs.~(\ref{fik}) and (\ref{fie}), we find
\begin{equation}
N_k = \frac{\sqrt{3}}{4r}\left(\sqrt{r\alpha(64+3r\alpha)} - r\sqrt{\alpha(4+3\alpha)}\right).
\label{Nk}
\end{equation}
A consistency relation can be derived by substituting Eq.~(\ref{fik}) into (\ref{Ins}), resulting in
\begin{equation}
\alpha = \frac{4r}{3\delta_{n_s}(4\delta_{n_s}-r)},
\label{cralfa}
\end{equation}
where $\delta_{n_s}\equiv 1-n_s$. While a consistency relation for $n_{sk}$ is
\begin{equation}
n_{sk} = -\frac{\delta_{n_s}^2}{2},
\label{crnsk}
\end{equation}
note that it does not involve the parameter $\alpha$.
By eliminating the parameter $\alpha$ from the expressions of interest and writing them solely in terms of the observables $n_s$ and $r$, as illustrated in Fig.~\ref{Nrebounds} for $N_{re}$, we can gain insights into the impact of the instant reheating condition ($N_{re}=0$) on the allowed ranges of $n_s$ and $r$.
This condition is analytically described by Eq.~(\ref{cir}). From Fig.~\ref{Nrebounds} we see that there is a curve in the $n_s$-$r$ plane for which $N_{re}=0$. To decide which point in that curve gives $N_{re}=0$ simultaneously with upper bounds for $N_k$ and $N_{rd}$ as expected (see Fig.~\ref{fd}), we can rewrite Eq.~(\ref{cir}) with the help of Eq.~(\ref{Treir1}) as follows
\begin{equation}
N_k=S_1+\frac{1}{4}\ln\left(\frac{135A_sr^2}{2g_{re}}      \left(\frac{T_{re}^{(ir)}}{Mpl}\right)^{-4}\right).
\label{cir2}
\end{equation}
For a given $T_{re}^{(ir)}$, $N_k$ is maximized by the upper bound on $r$. Alternatively, comparing with similar figures to Fig.~\ref{Nrebounds}, for $N_k$ and $N_{rd}$, we can quickly decide that the point $(n_s,r)=(0.9652,0.068)$ is the one that maximize both $N_k$ and $N_{rd}$ while keeping $N_{re}=0$. In the same way, from the Fig.~\ref{Nrebounds}, the point $(n_s,r)=(0.9578,0.068)$ maximize $N_{re}$ while minimizing both $N_k$ and $N_{rd}$  (see below, after Eq.~(\ref{nsasintotic})). Finally, using Eq.~(\ref{Nkeq}) we can check that our results are correct.

The upper bound for $N_k$ holds true only when $\omega_{re}<1/3$. If $\omega_{re}>1/3$, the upper bound for $N_k$ becomes a lower bound instead. This behavior can be visualized by examining the rhs dashed line in Fig.~\ref{fd}, representing the maximum number of radiation $e$-folds when $N_{re}=0$.

\begin{figure}[ht!]
\centering
\includegraphics[trim = 0mm  0mm 1mm 1mm, clip, width=12cm, height=8cm]{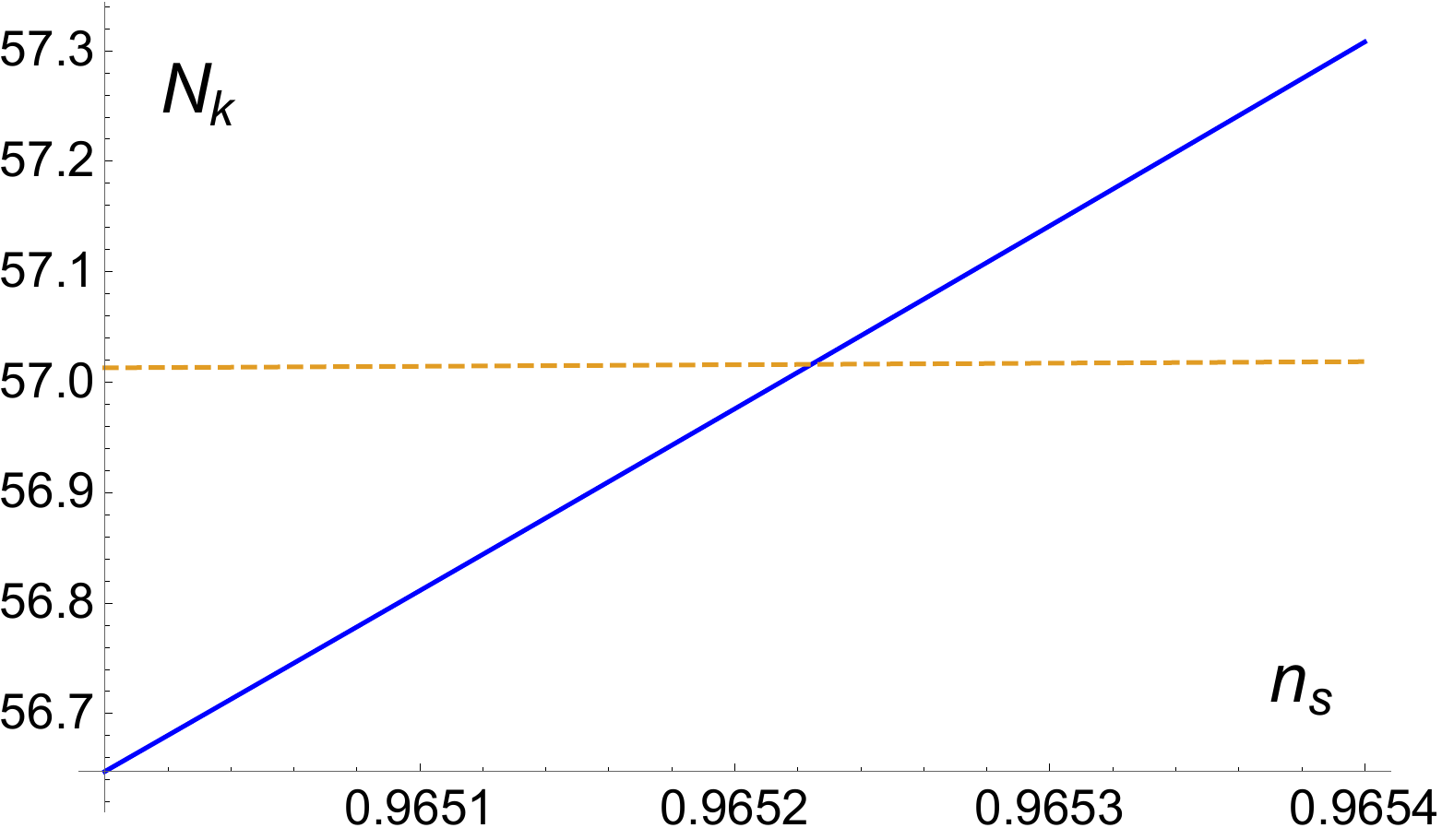}
\caption{Taking the upper bound for $r$ from Eq.~(\ref{boundcr}), the solid (blue) curve represents the lhs of Eq.~(\ref{cir}), using (\ref{Nk}) and (\ref{cralfa}), while the dashed (orange) curve represents the slowly varying function of the rhs of the same equation, where $f(\phi_k)$ and $f(\phi_e)$ are given by formulas (\ref{fk}) and (\ref{fe}), respectively. The point of intersection between these two curves provides us with the approximate value $n_s\approx 0.9652$, where the consistency relation for the $p=2$ model, expressed by Eq.~(\ref{cralfa}), has been used.}
\label{Nks}
\end{figure}

\begin{figure}[ht!]
\centering
\includegraphics[trim = 0mm  0mm 1mm 1mm, clip, width=12cm, height=8cm]{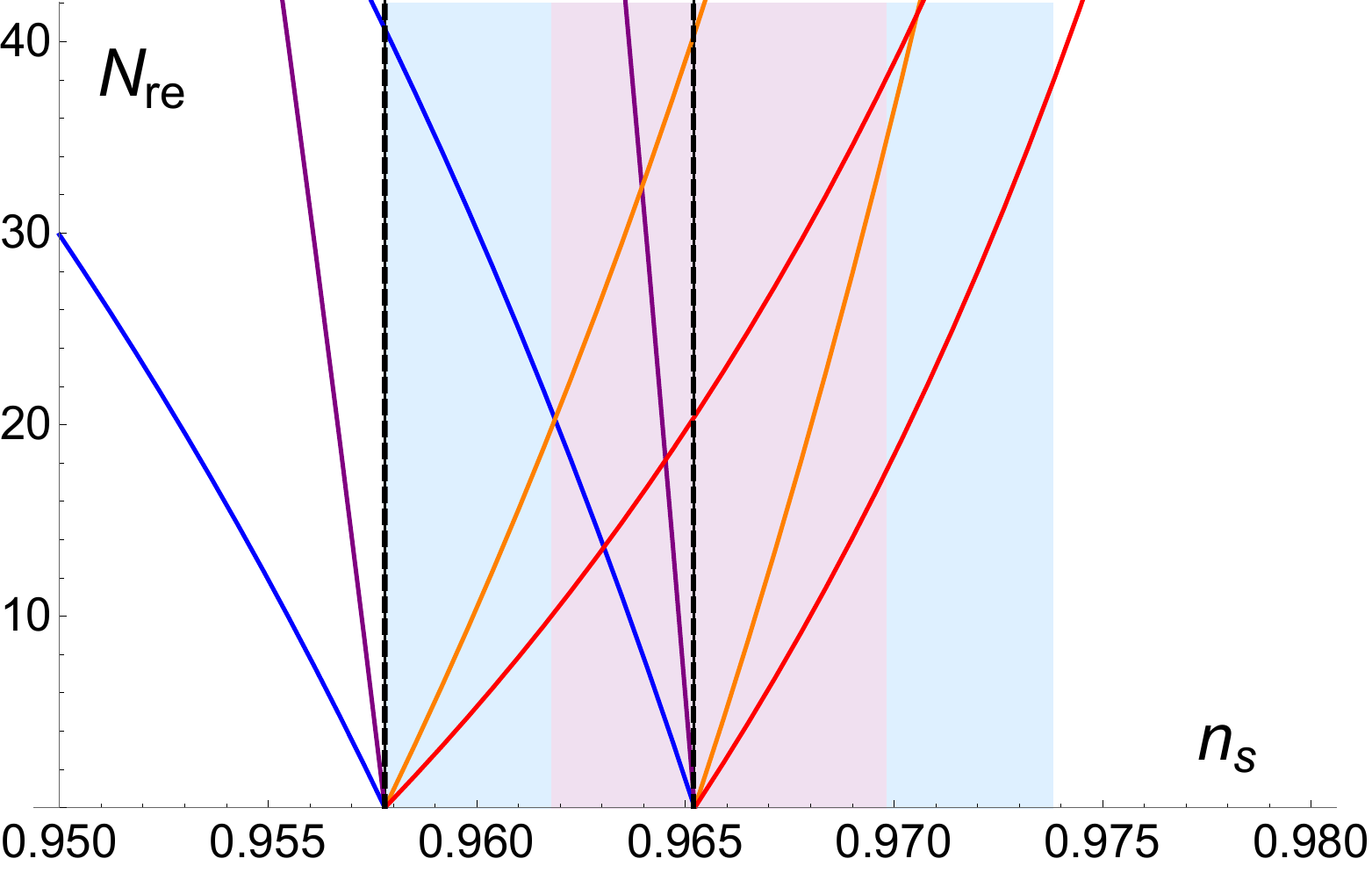}
\caption{The plot illustrates the number of $e$-folds of reheating $N_{re}$ as a function of the spectral index $n_s$, as given by Eq.~(\ref{Nre5}), for the model (\ref{potsech}) with $p=2$. From left to right, the curves correspond to different EoS for reheating: $\omega_{re}=0, \frac{1}{4}, \frac{2}{3}, 1$, respectively. In the rhs plot, the value of $\alpha$ is fixed at $\alpha=36.7$, while in the lhs plot, the value of $\alpha$ is set to $\alpha=1.4\times 10^{-14}$. The light blue (light purple) region corresponds to the $2\sigma$ ($1\sigma$) interval. The region bounded by the dashed vertical lines in this figure is defined by the constraints imposed by reheating considerations. Note that this figure (and Fig.~\ref{Treir}) provides only limited information compared to the more comprehensive insights offered by Fig.~\ref{Nrebounds}.  Although the figure is presented in the typical format found in the literature for this type of representation, one might question the relevance of plotting curves with $\omega_{re}\neq 0$ for an approximately quadratic potential around its minimum. One justification for doing so is that possible non-perturbative effects could lead to a non-zero constant effective equation of state (EoS) value.}
\label{Nreir}
\end{figure}

\begin{figure}[ht!]
\centering
\includegraphics[trim = 0mm  0mm 1mm 1mm, clip, width=12cm, height=8cm]{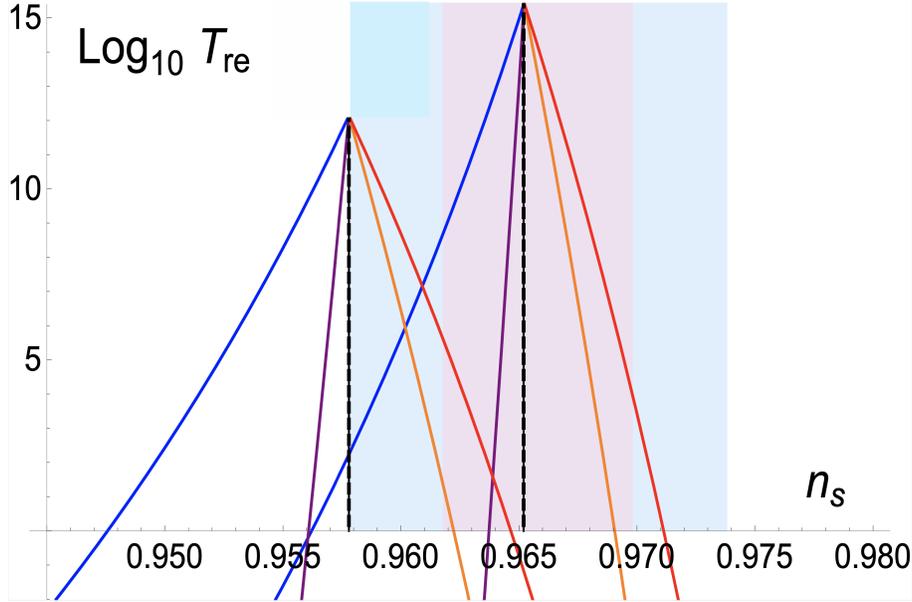}
\caption{The plot displays the reheating temperature $T_{re}$ as a function of the spectral index $n_s$, as given by Eq.~(\ref{Tre})), for the model (\ref{potsech}) with $p=2$. The curves from left to right correspond to different EoS for reheating: $\omega_{re}=0, \frac{1}{4}, \frac{2}{3}, 1$, respectively.
In the rhs plot, the value of $\alpha$ is fixed at $\alpha=36.7$, while in the lhs plot, the value of $\alpha$ is set to $\alpha=1.4\times 10^{-14}$. It is important to note that while instantaneous reheating always implies the same value for $N_{re}$ ($N_{re}=0$), the figure reveals that the value of the (instant) reheating temperature $T_{re}^{ir}$ from Eq.~(\ref{Treir}) depends on $n_s$. Specifically, it decreases as $n_s$ decreases. The light blue (light purple) region corresponds to the $2\sigma$ ($1\sigma$) interval. The region between the dashed vertical lines is determined by the bounds coming from the reheating constraints.  See comments at the end of figure caption \ref{Nreir}.}
\label{Treir}
\end{figure}

\begin{figure}[ht!]
\centering
\includegraphics[trim = 0mm  0mm 1mm 1mm, clip, width=12cm, height=8cm]{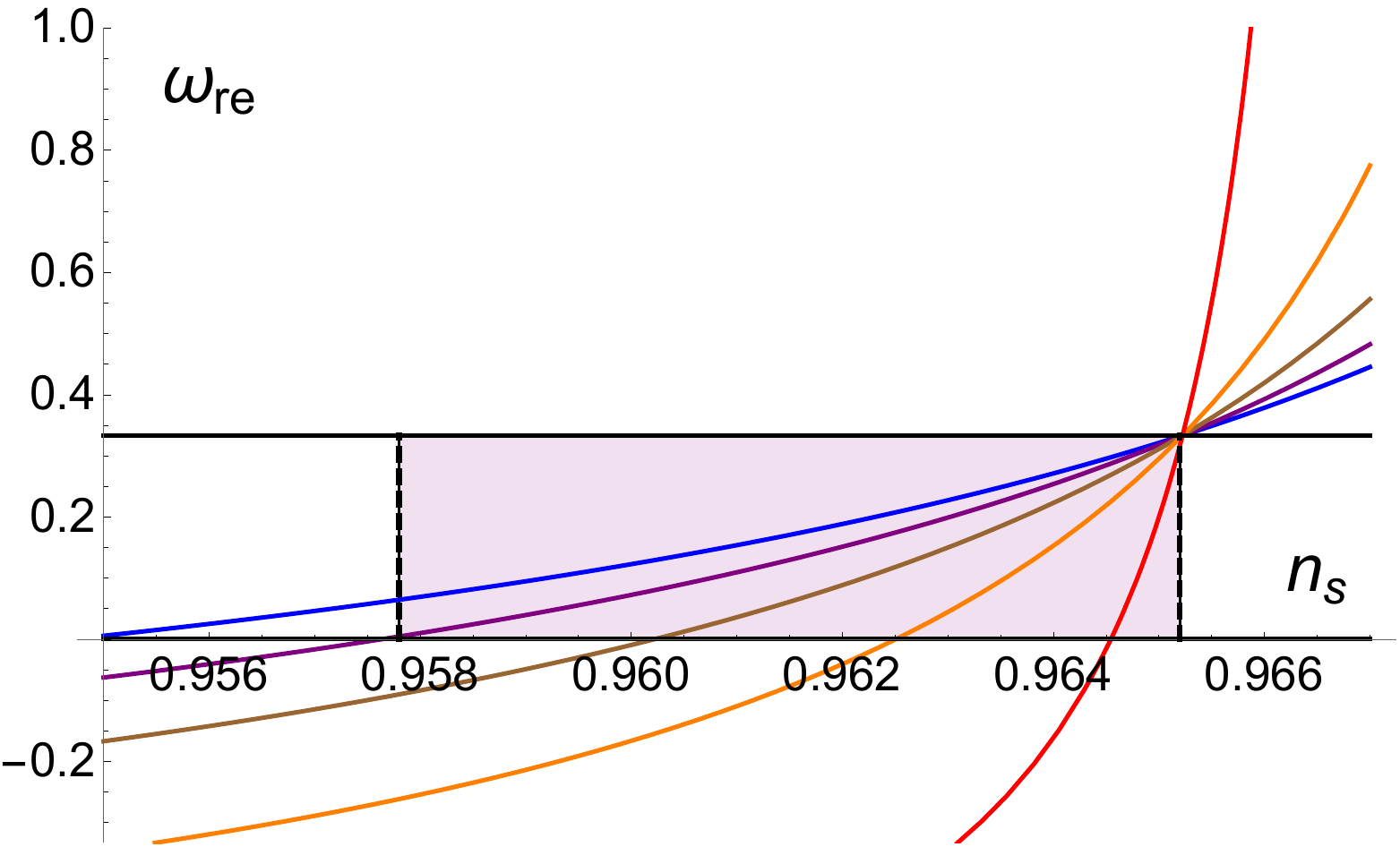}
\caption{The figure displays the equation of state parameter $\omega_{re}$, as given by Eq.~(\ref{wns}), as a function of the spectral index $n_s$ for different values of $T_{re}$, namely (from top to bottom) $T_{re}=10^{-2}, 10^{2}, 10^{6}, 10^{10}, 10^{14}$, while keeping $\alpha$ at its upper bound, $\alpha=36.7$. The consistency relation of the model is given by Eq.~(\ref{cralfa}). The shaded region corresponds to the allowed values of $n_s$ (see Table~\ref{tabla}).}
\label{wnsTre}
\end{figure}

\begin{figure}[ht!]
\centering
\includegraphics[trim = 0mm  0mm 1mm 1mm, clip, width=12cm, height=8cm]{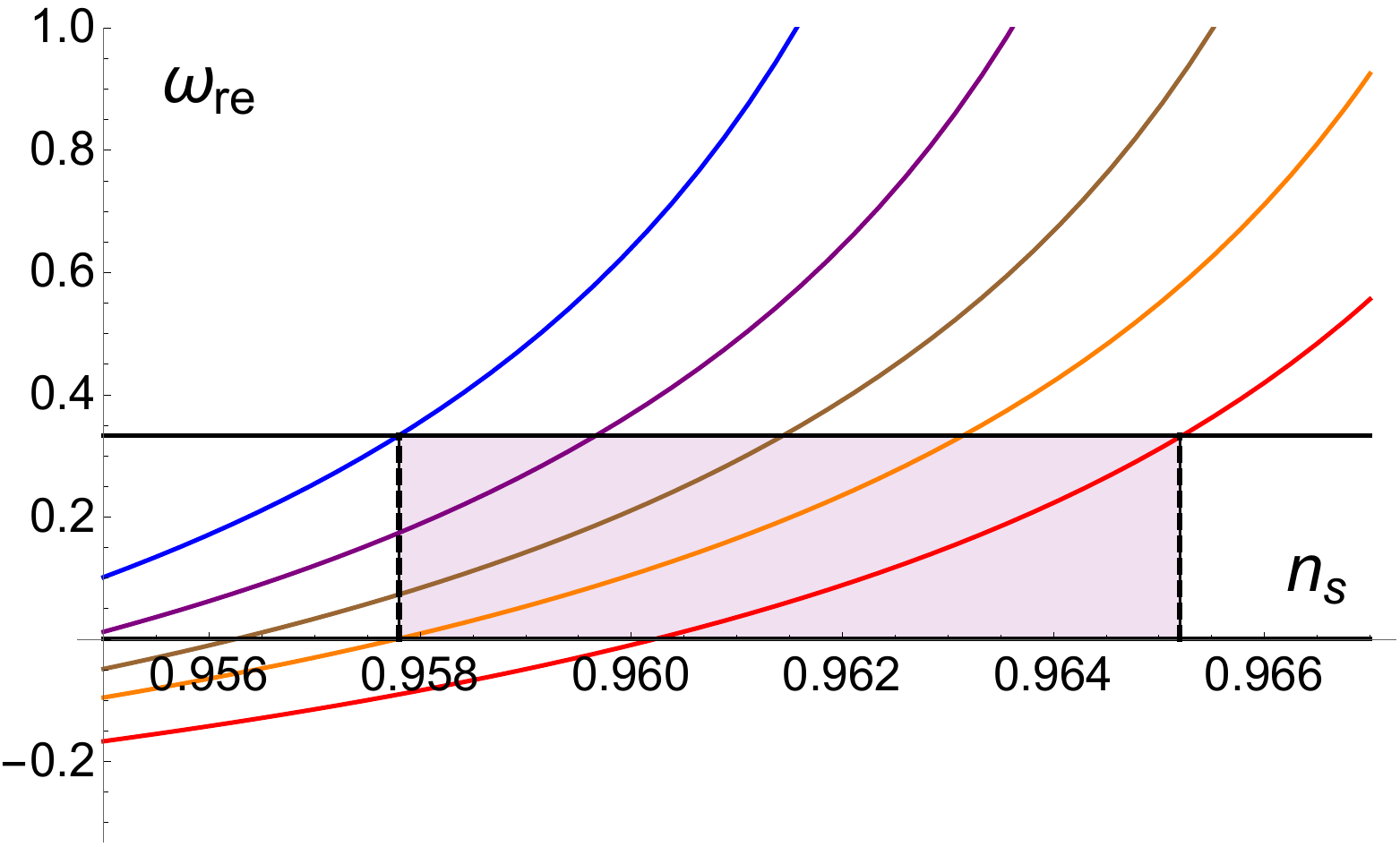}
\caption{The figure illustrates the equation of state parameter $\omega_{re}$ as a function of the spectral index $n_s$ for different values of $\alpha$ (from top to bottom): $\alpha=1.4\times 10^{-14}, 10^{-10}, 10^{-6}, 10^{-2}, 36.7$, while setting $T_{re}$ to an arbitrarily chosen value $T_{re}=10^6$\,GeV. The tensor-to-scalar ratio $r$ in  Eq.~(\ref{wns}) is related to $n_s$ and $\alpha$ through Eq.~(\ref{cralfa}).  The shaded region represents the allowed values of $n_s$ (see Table~\ref{tabla}).}
\label{wnsAlfa}
\end{figure}

The overall scale of the potential, denoted as $V_0$, can be expressed in terms of the observables $n_s$ and $r$ using Eq.~(\ref{IA})
\begin{equation}
V_0=\frac{3A_s \pi^2r}{2f(\phi_k)}M_{Pl}^4 = \frac{6A_s \pi^2 r\delta_{n_s}} {4\delta_{n_s}-r}M_{Pl}^4,
\label{V0}
\end{equation}
which reveals that $r<4\delta_{n_s}$. In the limiting case when $r=4\delta_{n_s}$, the potential takes the form of a quadratic monomial \cite{German:2021tqs}: from Eqs.~(\ref{cralfa}) and (\ref{V0}), the potential (\ref{potsech}) becomes
\begin{equation}
V = \frac{6A_s \pi^2 r\delta_{n_s}} {4\delta_{n_s}-r}\left(1-sech^2\left(\sqrt{\frac{(4\delta_{n_s}-r)\delta_{n_s}}{8r}} \frac{\phi}{M_{Pl}}\right)\right)M_{Pl}^4,
\label{potnsr}
\end{equation}
for $r\rightarrow 4\delta_{n_s}$
\begin{equation}
V =\frac{3}{64} A_s \pi^2 r^2\left(\frac{\phi}{M_{Pl}}\right)^2 M_{Pl}^4 + \cdot  \cdot  \cdot
\label{potlim}
\end{equation}
\begin{figure}[ht!]
\centering
\includegraphics[trim = 0mm  0mm 1mm 1mm, clip,  width=12cm, height=8cm]{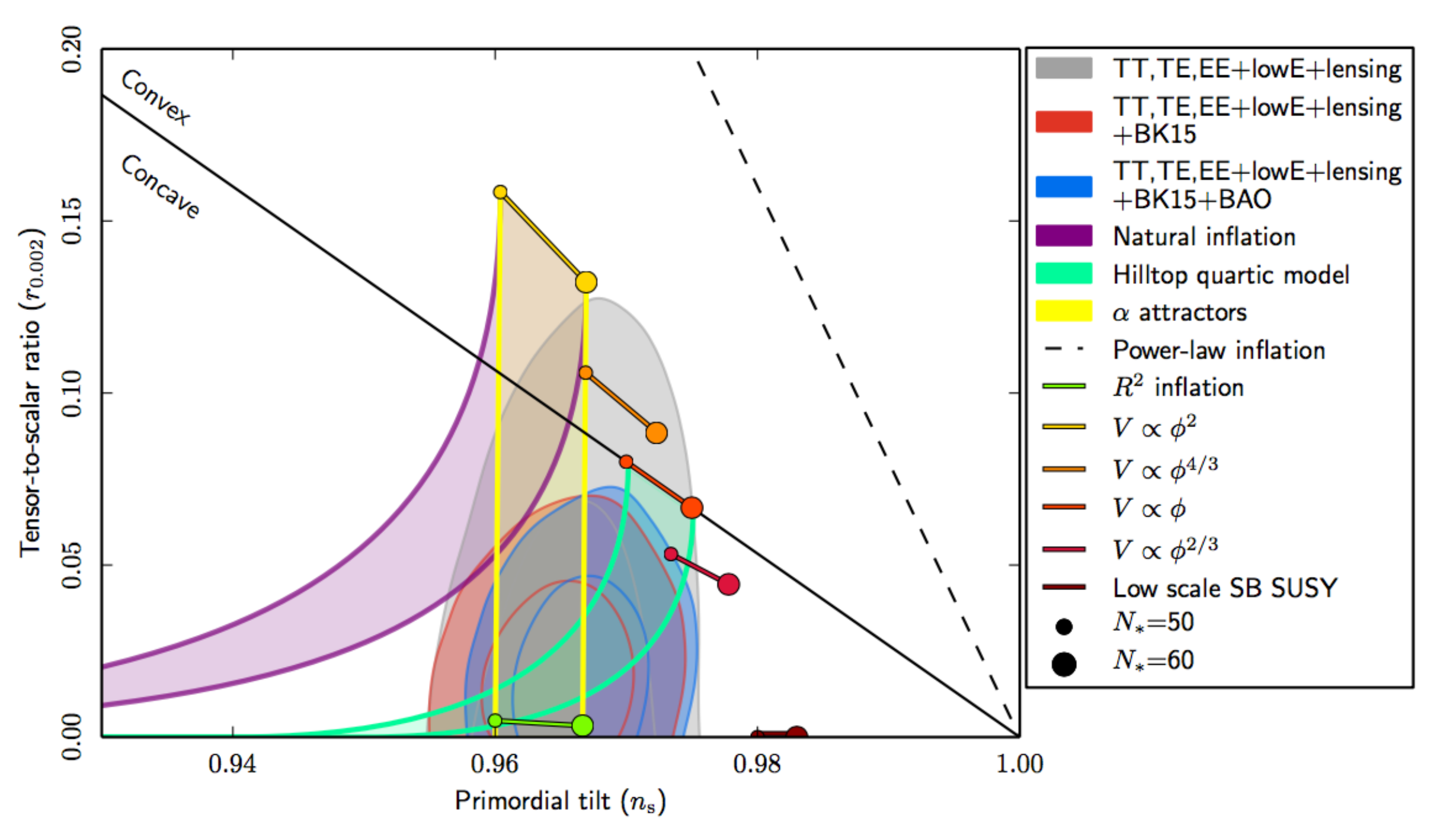}
\caption{Fig. 8 in the article by the Planck Collaboration 2018 \cite{Akrami:2018odb} displays the predictions of  monomial potentials and various other inflationary models (as described in the right-hand panel of the figure). Notably, the figure shows a substantial overlap between the predictions of $\alpha$-attractor inflation (indicated by the yellow curves, which are almost vertical for the $p=2$ case) and the Planck data alone, as well as when combined with the BICEP2/Keck Array (BK15) \cite{BICEP2:2015ns} or BICEP2/Keck Array+Baryon Acoustic Oscillations (BK15+BAO) data.}
\label{Pl}
\end{figure}
Thus, the fact that $\alpha$-attractor models end in monomials (see Fig.~\ref{Pl}) is easy to understand. We can also understand the rapid rise of the yellow curves in the Fig.~\ref{Pl} by writing $N_k$ entirely in term of the observables by elliminating the parameter $\alpha$ with Eq.~(\ref{cralfa})
\begin{equation}
N_{k} = \frac{8\delta_{n_s}-r-\sqrt{r^2+r\delta_{n_s}(4\delta_{n_s}-r)}}{\delta_{n_s}(4\delta_{n_s}-r)}\;.
\label{Nkexplicit}
\end{equation}
Solving for $r$
\begin{equation}
r = \frac{4(N_{k}\delta_{n_s} -2)^2}{1+N_{ke} (N_{k} \delta_{n_s}-2)}\;.
\label{rdensNkep2}
\end{equation}
This solution should be supplemented with the condition
\begin{equation}
1-\frac{2}{N_{k}} < n_s < 1-\frac{2}{N_{k}}+\frac{1}{N_{k}^2}\;,
\label{conditions}
\end{equation}
which guarantee that $r$ is a well defined real positive number. The rhs bound is larger than the lhs bound by a quadratic term in $1/N_{k}$, explaining the rapid raise of the yellow curves in Fig.~\ref{Pl}. It also shows the deviation from the commonly quoted asymptotic result
\begin{equation}
n_s=1-\frac{2}{N_{k}}+\cdot \cdot \cdot\, .
\label{nsasintotic}
\end{equation}
As stated before, by taking the upper bound for $r$ from Eq.~(\ref{boundcr}), we can directly solve Eq.~(\ref{cir}) for $n_s$, as shown in Fig.~\ref{Nks}. The solution yields $n_s\approx 0.9652$ and $\alpha\approx 36.75$. From these values, we can deduce that $T_{re}^{(ir)}\approx 2.7\times 10^{15}$ GeV and $N_k\approx 57.0$. It is evident that the lower bound for the number of $e$-folds of reheating is given by $N_{re}=0$, and using Eq.~(\ref{Nrd1}), we obtain $N_{rd}\approx 57.6$. 
We observe that the constraint on $n_s$ obtained directly from our analysis is considerably stronger than the constraint derived from Eq.~(\ref{boundsns}), which, at the $2\sigma$ confidence level, yields a value of 0.9738 (see Figs.~\ref{Nreir} and \ref{Treir}). The scale of inflation, denoted as $\Delta$ and defined as $\Delta\equiv V(\phi_k)^{1/4}$, can be determined using Eq.(\ref{IA}), resulting in $\Delta \approx 1.7\times 10^{16}$GeV while $V_0^{1/4}$, from Eq.~(\ref{cir}), is $V_0^{1/4}\approx 2.0\times 10^{15}$GeV, that is, $V_0^{1/4}$ is $18\%$ larger than $\Delta$ showing the effect of the $f(\phi_k)$-term on scale invariance. These findings and other related results are summarized in the Table \ref{tabla}.
We can establish lower bounds for $\alpha$ and $r$ by considering the lower limit for $n_s$ at $2\sigma$, which is $0.9578$. Using the consistency relation (\ref{cralfa}), we can eliminate $r$ from Eq.~(\ref{cir}) and solve for $\alpha$ giving $\alpha\approx 1.4\times 10^{-14}$. From here we deduce that   $r\approx 7.5\times 10^{-17}$. Additionally, we obtain $T_{re}^{(ir)}\approx 1.4\times 10^{12}$ GeV.  
To determine an upper bound for $N_{re}$, we refer to Fig.~\ref{Nrebounds}. From this figure, we observe that the maximum value of $N_{re}$ occurs at the point $(r,n_s)=(0.068,0.9578)$. Therefore, using Eq.(\ref{Nre5}), we obtain $N_{re}\approx 40.1$. 
Using Eq.~(\ref{Nrd1}), we find $N_{rd}\approx 27.6$, $N_k\approx 46.9$, an inflationary scale of $\Delta \approx 1.6\times 10^{16}$ GeV, and $V_0^{1/4}\approx  2.0 \times 10^{16}$ GeV. It is worth noting that while the value of $N_{re}$ for instant reheating is always $N_{re}=0$ (as depicted in Fig.~\ref{Nreir}), the value of $T_{re}^{(ir)}$ depends on $r$ (specifically, $T_{re}^{(ir)}\propto r^{1/4}$ for small $r$), this is illustrated in Fig.~\ref{Treir}. Finally, Fig.~\ref{wnsTre} displays the equation of state parameter $\omega_{re}$, as given by Eq.~(\ref{wns}), as a function of the spectral index $n_s$ for different values of $T_{re}$, while Fig.~\ref{wnsAlfa} illustrates $\omega_{re}$ as a function of $n_s$ for different values of $\alpha$.
These results, along with additional findings, are summarized in the Table \ref{tabla}

\begin{table*}[htbp!]
 \begin{center}
{\begin{tabular}{cccc}
\small
Characteristic & Range & Characteristic& Range  \\
\hline \hline
{$k_p $} & $0.05Mpc^{-1}$ & $M_{Pl}$ & $2.43568\times 10^{18}\mathrm{GeV}$ \\
{$T_0 $}   & $2.7255\,K$ & $g_{re}$ & $106.75$  \\
{$A_s $}  & $2.1\times 10^{-9}$ &  $g_{s,re}$ & $106.75$  \\
 \hline \hline
Characteristic & Range & Characteristic& Range  \\
\hline \hline
{$n_{s}$} & $\left(0.9578,0.9652\right)$ & $N_k$ & $\left(46.9,57.0\right)$  \\
{$r$} & $\left(0.068,7.5 \times 10^{-17}\right)$ & $N_{re}$ & $\left(40.1,0\right)$   \\
{$n_{t}$} & $\left(-8.5\times 10^{-3},-9.4 \times 10^{-18}\right)$ &  $N_{rd}$ & $\left(27.6,57.6\right)$\\
{$n_{sk}$} &  $\left(-8.9 \times 10^{-4},-6.0\times 10^{-4}\right)$ & $N_k+N_{re}+N_{rd}$ & $114.6$  \\
{$\alpha$} & $\left(36.7,1.4 \times 10^{-14}\right)$ & $N_{keq}$ & $114.6$  \\
{$V_0^{1/4}$/GeV } & $\left(2.0\times 10^{16}\,,3.0\times 10^{12}\right)$ & $T_{re}$/GeV & $\left(263,2.7\times 10^{15}\right)$  \\\hline\hline
\end{tabular}}
\caption{We provide the parameter values used in our calculations and present the obtained values for the characteristics of the model. These values are derived by considering the constraints provided in the Table 3 of the article \cite{Akrami:2018odb} for the cosmological model $\Lambda$CDM$+r+dn_s/d\ln k$ with the data set Planck TT,TE,EE+lowE+lensing+BK15+BAO. These bounds offer constraints on the parameters and observables within this specific cosmological model and data combination.
To determine the values of the observables, we solve Eq.(\ref{cir}) for the spectral index $n_s$. By using the consistency relations given by Eqs.(\ref{Int}) and (\ref{crnsk}), we obtain the remaining observables. The number of $e$-folds during inflation $N_k$, reheating $N_{re}$, and radiation $N_{rd}$ are calculated using Eqs.~(\ref{Nk}), (\ref{Nre5}), and (\ref{Nrd1}), respectively. Additionally, the sum of these three quantities equals $N_{keq}$ as given by Eq.~(\ref{Nkeq}), which solely depends on the parameter $r$. For further details, refer to the paragraph following Eq.~(\ref{Nkeq}). Quantities are written to indicate correspondence between them e.g., to $T_{re}=263$\,GeV corresponds $N_{re}=40.1$, $N_k=46.9$, and so on.}
\label{tabla}
\end{center}
\end{table*}
\newpage

\subsection {\bf The case $p=2$ without running}\label{caso2} 
The case $p=2$ without running is based in the analysis of the following bounds (see the TABLE I of \cite{Paoletti:2022anb})
\begin{equation}
n_s =0.9653\pm 0.0041    \quad    (68\%\,\, C.L.),
\label{boundsnsnr} 
\end{equation}
\begin{equation}
r < 0.035     \quad    (95\%\,\, C.L.).
\label{boundsrnr} 
\end{equation}
As in the previous case we extend the range of $n_s$ up to $2\sigma$ and proceed in a similar way as before. The results are displayed in the  Table \ref{tablanr}.
\begin{table*}[htbp!]
 \begin{center}
{\begin{tabular}{cccc}
\small
Characteristic & Range & Characteristic& Range  \\
\hline \hline
{$n_{s}$} & $\left(0.9571,0.9650\right)$ & $N_k$ & $\left(46.1,56.7\right)$  \\
{$r$} & $\left(3.4 \times 10^{-18},0.035\right)$ & $N_{re}$ & $\left(41.8,0\right)$   \\
{$n_{t}$} & $\left(-4.2 \times 10^{-19},-4.4\times 10^{-3}\right)$&  $N_{rd}$ & $\left(26.3,57.6\right)$\\
 {$\alpha$} & $\left(6.1 \times 10^{-16},12.7\right)$  & $N_k+N_{re}+N_{rd}$ & $114.263$  \\
 {$V_0^{1/4}$/GeV } & $\left(1.4\times 10^{12},1.5\times 10^{16}\right)$  & $N_{keq}$ & $114.263$  \\
$T_{re}$/GeV & $\left(71.4,2.7\times 10^{15}\right)$ & \\
\hline\hline
\end{tabular}}
\caption{The bounds considered here are derived from the TABLE I of the article \cite{Paoletti:2022anb} with Planck+BK18 data and are provided explicitly in Eqs.(\ref{boundsnsnr}) and (\ref{boundsrnr}) for the cosmological model $\Lambda$CDM$+r$. For a more detailed explanation of how these bounds were derived, please refer to the caption of Table \ref{tabla}}
\label{tablanr}
\end{center}
\end{table*}

\subsection {\bf Dependence of cosmological quantities  on the parameter $\alpha$}\label{caso3} 

We can explore the dependence of the quantities studied earlier, for the $p=2$ $\alpha$-attractor model, on the parameter $\alpha$. Our starting point is Eq.~(\ref{Nre5}), which we write here separating the part that depends on the model of inflation
\begin{equation}
N_{re}=\frac{1}{1-3\,\omega_{re}}\ln\left( \frac{\left(\frac{43}{11g_{s,re}}\right)^{4/3}g_{re}A_s\pi^4T_0^4}{270k_p^4}\right)+\frac{1}{1-3\,\omega_{re}}\ln\left( \frac{e^{-4N_k}f(\phi_k)r}{f(\phi_e)}\right),
\label{Nre6}
\end{equation}
where the first term is just a number.  In terms of the potential function we find $f(\phi_k)=1-sech^2\left(\frac{\phi_k}{\sqrt{6\alpha}M_{Pl}}\right)$. We can eliminate $r$ from  Eq.~(\ref{Nre6}) by rewriting it as $r=16\epsilon_k$, where $\epsilon_k$ represents the SR parameter $\epsilon$ at horizon crossing, $\phi=\phi_k$, we obtain 
\begin{equation}
r=\frac{16\left(1-f(\phi_k)\right)^2}{3\,\alpha f(\phi_k)}.
\label{r}
\end{equation}
 From Eq.~(\ref{Nk0}) we can eliminate $\phi_k$ in terms of $N_k$ and the parameter $\alpha$ to get
\begin{equation}
f(\phi_k)=1-\frac{3\alpha}{2N_k+\frac{3\alpha}{1-f(\phi_e)}},
\label{fikNkalfa}
\end{equation}
where the value of $f(\phi_e)$ is given  by Eq.~(\ref{fe}). For the $p=2$ $\alpha$-attractor model, it can be shown that $N_{re}$ can be expressed  as
\begin{equation}
N_{re}=S_2+\frac{1}{1-3\,\omega_{re}}\ln\left(\frac{\alpha\, e^{-4N_k}}{\left(2N_k+\frac{3\alpha}{1-f(\phi_e)}\right)^2f(\phi_e)}\right),
\label{NreNkalfa}
\end{equation}
where we have left $\omega_{re}$ unspecified and
\begin{equation}
S_2 =\frac{1}{1-3\,\omega_{re}}\ln\left(\frac{8\left(\frac{43}{11g_{s,re}}\right)^{4/3}g_{re}\pi^4 A_s T_0^4 }{45k_p^4 }\right)\approx \frac{230.6}{1-3\,\omega_{re}}.
\label{S2}
\end{equation}
\begin{figure}[ht!]
\centering
\includegraphics[trim = 0mm  0mm 1mm 1mm, clip, width=12cm, height=8cm]{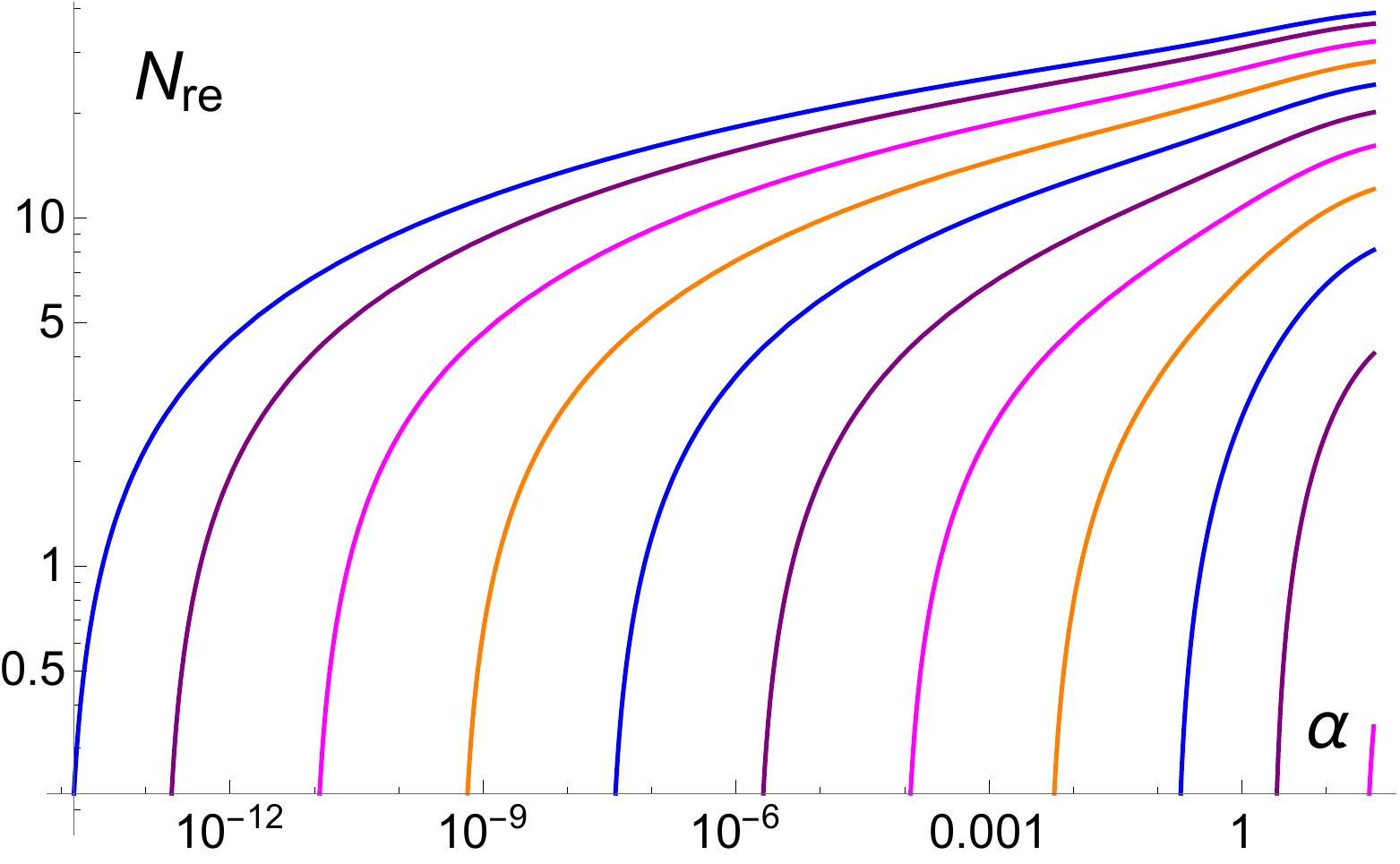}
\caption{The plot illustrates the relationship between the number of $e$-folds during reheating, denoted as $N_{re}$ and given by Eq.~(\ref{NreNkalfa}), and the parameter $\alpha$, considering various values of $N_k$ and $\omega_{re}=0$. While Eq.~(\ref{Nre6}) provides a general formula relating reheating and inflation applicable to any inflation model, Eq.~(\ref{NreNkalfa}) specifically corresponds to the $p=2$ $\alpha$-attractor model previously studied. Comparing this plot with Fig.~\ref{Nrebounds}, we observe that by generating an infinite number of curves as described, we would cover the entire orange surface depicted in Fig.~\ref{Nrebounds}. Thus, Eq.~(\ref{NreNkalfa}) allows for the construction of individual trajectories of $N_{re}$ parametrized by $\alpha$.}
\label{Nrealfa}
\end{figure}
A plot of $N_{re}$ as a function of $\alpha$ for several values of $N_k$ is shown in Fig.~\ref{Nrealfa}.

\begin{figure}[ht!]
\centering
\includegraphics[trim = 0mm  0mm 1mm 1mm, clip, width=12cm, height=8cm]{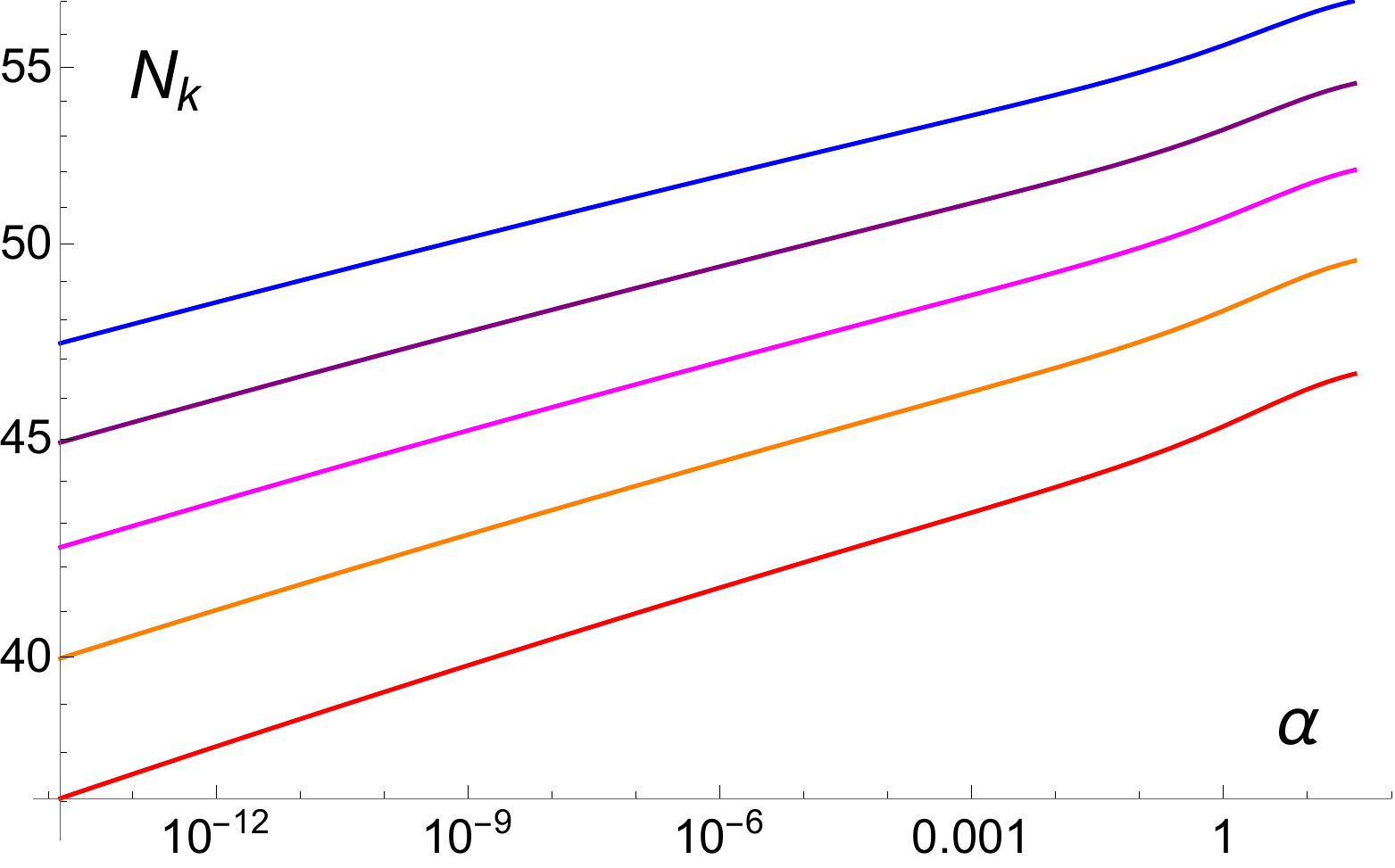}
\caption{The plot illustrates the relationship between the number of $e$-folds during inflation, denoted as $N_k$, and the parameter $\alpha$ for various values of $N_{re}$ and a fixed value of $\omega_{re}=0$. The considered values of $N_{re}$ range from top to bottom: 0, 10, 20, 30, and 40. Moreover, by examining the endpoints of the upper curve in the plot, one can derive the bounds for $N_k$ presented in Tables~\ref{tabla} and \ref{tablanr}.}
\label{Nkalfa}
\end{figure}

We can derive an expression for $N_k$ in terms of $N_{re}$ and $\alpha$ by inverting Eq.~(\ref{NreNkalfa})
\begin{equation}
N_k=-\frac{3\alpha}{2(1-f(\phi_e))}+\frac{1}{2}ProductLog\left[\frac{\sqrt{\alpha}\, e^{\frac{3\alpha}{1-f(\phi_e)}}}{e^{\frac{1}{2}(1-3\omega_{re})(N_{re}-S_2)f^{1/2}(\phi_e)}}\right],
\label{NkNrealfa}
\end{equation}
where the function $ProductLog$ is defined as used by Mathematica: a dedicated implementation of the Lambert $W$ function, designed specifically for the principal branch of the function. A plot of $N_k$ as a function of $\alpha$ for several values of $N_{re}$ is shown in Fig.~\ref{Nkalfa}.

We can solve Eq.~(\ref{Nk}) for $r = r (N_k,\alpha)$ or, using Eq.~(\ref{cralfa}), obtain $n_s = n_s (N_k, \alpha)$ and so on, where $N_k$ is given by Eq.~(\ref{NkNrealfa}). By following this procedure, we can express all relevant cosmological quantities in terms of $N_{re}$ and $\alpha$ (see \cite{Iacconi:2023mnw} for more details on this approach).
\subsection{\bf Further considerations on the reheating temperature}\label{caso4}

Up to this point, we have considered the model defined by equation (\ref{potsech}) without any consideration about its origin. However, in gravitational theories, including supersymmetric theories and those associated with $\alpha$-attractors, the gravitino problem is a frequent issue \cite{Ellis:1982yb}. To avoid this problem, it is crucial to restrict the reheating temperature to less than $10^9$ GeV in order to effectively prevent the overproduction of gravitinos and other relic fields. Failing to do so could pose challenges to the success of Big Bang nucleosynthesis \cite{Kawasaki:2006gs},  \cite{Kawasaki:2006hm},  \cite{Kohri:2005wn}.

Proceeding similarly to section \ref{caso1}, it is straightforward to demonstrate that imposing the limit of $10^9$ GeV for the reheating temperature leads to obtaining new bounds that would replace those presented in Table~\ref{tabla}. These results are presented in Table~\ref{tabla109}.
\begin{table*}[htbp!]
 \begin{center}
{\begin{tabular}{cccc}
\small
Characteristic & Range & Characteristic& Range  \\
\hline \hline
{$n_{s}$} & $\left(0.9578,0.9619\right)$ & $N_k$ & $\left(46.9,52.0\right)$  \\
{$r$} & $\left(0.068,1.4 \times 10^{-10}\right)$ & $N_{re}$ & $\left(40.1,19.8\right)$   \\
{$n_{t}$} & $\left(-8.5\times 10^{-3},-1.7\times 10^{-11}\right)$ &  $N_{rd}$ & $\left(27.6,42.8\right)$\\
{$n_{sk}$} &  $\left(-8.9 \times 10^{-4},-7.2\times 10^{-4}\right)$ & $N_k+N_{re}+N_{rd}$ & $114.6$  \\
{$\alpha$} & $\left(21.3,2.6 \times 10^{-8}\right)$ & $N_{keq}$ & $114.6$  \\
{$V_0^{1/4}$/GeV } & $\left(1.9\times 10^{16}\,,1.1\times 10^{14}\right)$ & $T_{re}$/GeV & $\left(263,10^{9}\right)$ \\\hline\hline
\end{tabular}}
\caption{We provide the parameter values employed in our computations and present the obtained values for the model characteristics. These values are determined while ensuring that $T_{re}<10^9$\,GeV. We derive these values by taking into account the limits specified in Table 3 of  \cite{Akrami:2018odb}, which pertain to the cosmological model $\Lambda$CDM$+r+dn_s/d\ln k$ with the Planck TT,TE,EE+lowE+lensing+BK15+BAO dataset. These restrictions impose constraints on the parameters and observables within this particular cosmological model and combination of data.
To ascertain the values of the observables, we solve Eq.~(\ref{cir}) for the spectral index $n_s$. By using the consistency relations provided by Eqs.~(\ref{Int}) and (\ref{crnsk}), we obtain the remaining observables. The number of $e$-folds during inflation $N_k$, reheating $N_{re}$, and radiation $N_{rd}$ are computed employing Eqs.~(\ref{Nk}), (\ref{Nre5}), and (\ref{Nrd1}), respectively. Furthermore, the sum of these three quantities is equivalent to $N_{keq}$ as expressed in Eq.(\ref{Nkeq}), which solely relies on the parameter $r$. For more detailed information, refer to the paragraph  following to Eq.~(\ref{Nkeq}). Quantities are written to indicate correspondence between them e.g., to $T_{re}=10^9$\,GeV corresponds $N_{re}=19.8$, $N_k=52.0$, and so on. }
\label{tabla109}
\end{center}
\end{table*}

\section{Conclusions}\label{con}

Inflationary models provide a framework for understanding the dynamics of the early universe. The values of the reheating temperature $T_{re}$ and the duration of reheating $N_{re}$ are not only important in their own right but also have implications for other observables. Consistency relations among observables, such as the spectral index $n_s$, and the tensor-to-scalar ratio $r$, offer additional constraints on inflationary models. These relations arise from the underlying dynamics of the inflationary field and provide valuable insights into the internal consistency of cosmological models. The generalized $\alpha$-attractor models of inflation have attracted significant attention due to their ability to accurately reproduce observed quantities. These models, inspired by concepts in supergravity and string theory, exhibit attractor behavior and offer a broader range of initial conditions that lead to similar observational outcomes. To investigate the aforementioned aspects, we focus on the study of reheating conditions, establishing consistency relations among observables, and exploring specific cases within the generalized $\alpha$-attractor model. We delve into the conditions for instantaneous reheating, deriving equations that allow us to determine the values of the spectral index and tensor-to-scalar ratio for which $N_{re}=0$. We also investigate the maximum number of $e$-folds during the inflationary and radiation eras. Additionally, we establish the consistency relations among observables and derive a formula for the instantaneous reheating temperature. We introduce a generalization of the basic $\alpha$-attractor model given by the potential $V =V_0\left(1-sech^{p}\left(\frac{\phi}{\sqrt{6\alpha}M_{Pl}}\right)\right)$, that ensures a positive-definite potential. This modification allows for the inclusion of odd and fractional values of the power $p$, expanding the range of viable models. Notably, this class of potentials exhibits quadratic behavior around its minimum, regardless of the value of $p$. As an example of how to proceed, we examine the implications of this generalized model in two specific cases: the case of $p = 2$, including the running of the spectral index, within the cosmological model $\Lambda$CDM$+r+dn_s/d\ln k$. We use data from collaborations such as Planck TT, TE, EE + lowE + lensing + BK15 + BAO to constrain the model and compare its predictions to observational data. We also study the same case of $p = 2$, but without considering the running of the spectral index, within the cosmological model $\Lambda$CDM$+r$. We use data from the Planck and BICEP/Keck 2018 collaborations to further constrain the model.
The derived constraints on $T_{re}$, $N_{re}$, $n_s$, and $r$ provide valuable information for assessing the validity of inflationary models and can be used to specify priors in Bayesian analyses of specific models.

\section*{Acknowledgments}

{The authors  acknowledge support from program UNAM-PAPIIT, grants IN107521 “Sector Oscuro y Agujeros Negros Primordiales” and IG102123 ``Laboratorio de Modelos y Datos (LAMOD) para proyectos de Investigaci\'on Cient\'ifica: Censos Astrof\'isicos".  L. E. P. and J. C. H. acknowledge sponsorship from CONAHCyT Network Project No.~304001 ``Estudio de campos escalares con aplicaciones en cosmolog\'ia y astrof\'isica'', and through grant CB-2016-282569.  The work of L. E. P. is also supported by the DGAPA-UNAM postdoctoral grants program, by CONAHCyT M\'exico under grants  A1-S-8742, 376127 and FORDECYT-PRONACES grant No. 490769.}
\\
\\
Data Availability Statement: No Data associated in the manuscript.

\end{document}